\documentclass[conference]{IEEEtran}

\usepackage{microtype}
\usepackage{graphicx}
\usepackage{amsfonts}
\usepackage{booktabs}
\usepackage{caption}
\usepackage{subcaption}
\usepackage{hyperref}
\usepackage{algorithm, algorithmic}
\usepackage{multirow}
\usepackage{verbatim}

\usepackage{amsmath,bm}
\newcommand{\z}{{\bm z}}
\newcommand{\x}{{\bm x}}
\newcommand{\m}{{\bm \mu}}
\newcommand{\si}{{\bm \sigma}}
\newcommand{\enc}{{\textbf{Enc}}}
\newcommand{\dec}{{\textbf{Dec}}}
\newcommand{\post}{{\textbf{Post}}}
\newcommand{\norm}[1]{\left\lVert#1\right\rVert}

\usepackage{ulem}
\usepackage{xcolor}

\begin{document}
\title{Enhancing Zero-Shot Many to Many Voice Conversion via Self-Attention VAE with Structurally Regularized Layers
}

\author{\IEEEauthorblockN{Ziang Long}
\IEEEauthorblockA{\textit{Department of Mathematics} \\
\textit{University of California}\\
Irvine, CA, USA \\
zlong6@uci.edu}
\and
\IEEEauthorblockN{Yunling Zheng}
\IEEEauthorblockA{\textit{Department of Mathematics} \\
\textit{University of California}\\
Irvine, CA, USA \\
yunliz1@uci.edu}
\and
\IEEEauthorblockN{Meng Yu}
\IEEEauthorblockA{\textit{Tencent AI Lab} \\
\textit{Tencent at Bellevue}\\
Seattle, WA, USA \\
raymondmyu@tencent.com}
\and
\IEEEauthorblockN{Jack Xin}
\IEEEauthorblockA{\textit{Department of Mathematics} \\
\textit{University of California}\\
Irvine, CA, USA \\
jack.xin@uci.edu}
}
\maketitle
\begin{abstract}
Variational auto-encoder (VAE) is an effective neural network architecture to disentangle a speech utterance into speaker identity and linguistic content latent embeddings, then generate an utterance for a target speaker from that of a source speaker. 
This is possible by concatenating the identity embedding of the target speaker and the content embedding of the source speaker uttering a  desired sentence. 
In this work, we propose to improve VAE models with self-attention and structural regularization (RGSM). 
Specifically, we found a suitable location of VAE's decoder to add a self-attention layer for incorporating non-local information in generating a converted utterance and hiding the source speaker's identity. We applied relaxed group-wise splitting method (RGSM) to regularize network weights and remarkably enhance generalization performance.

In experiments of zero-shot many-to-many voice conversion task on VCTK data set, with the self-attention layer and relaxed group-wise splitting method, our model achieves a gain of speaker classification accuracy on unseen speakers by 28.3\% while slightly improved conversion voice quality in terms of MOSNet scores. 
Our encouraging findings point to future research on integrating more variety of attention structures in VAE framework while controlling model size and overfitting for advancing zero-shot many-to-many voice conversions.
\footnote{The work was partially supported by NSF grants DMS-1854434 and DMS-1952644 at UC Irvine.}

\end{abstract}

\section{Introduction}
VC(voice conversion) is to convert speech of a source speaker to that of a target speaker while preserving its linguistic content. 
Recent works \cite{parrotron,huang2019voice} are able to do high quality End-to-End VC with parallel data, i.e. speech pairs of two speakers pronounce the same sentences. Yet, parallel training has the following obstacles: i) parallel data are expensive to collect, and utterance alignment takes even more effort, ii) VC to a different target requires re-training, iii) unable to do zero shot VC, i.e. conversion from/to the voice of an unseen speaker with only a few of his/her utterances. 

To overcome these limitations above, non-parallel voice conversion models are proposed. Several studies use extra data \cite{lee2006map, saito2011one} or additional models  \cite{xie2016kl} to facilitate training process, although it brings more cost of training. To avoid such disadvantages, recent studies introduced deep generative models, such as GANs \cite{hsu2017voice, kaneko2017parallel}, and VAEs \cite{hsu2017voice, kameoka2019acvae}.
Among them, CycleGAN-VC \cite{kaneko2018cyclegan} (and the enhanced version \cite{kaneko2020cyclegan}) is a GAN based model by configuring CycleGAN model (as widely used in image style transfer) with a gated CNN and cycle consistency loss. It produced  promising results on parallel-VC without parallel corpus for training. However, it is only designed for one-to-one voice conversion. In contrast, StarGAN-VC \cite{kameoka2018stargan} and its variants \cite{kaneko2019stargan, zhang2020gazev} provided many-to-many voice conversion on non-parallel corpus by only single generator. Adaptations from StarGAN \cite{choi2018stargan} include embedding loss and source-and-target conditional adversarial loss to enhance the models' generation accuracy.

However, the GAN based methods encounter saddle point problem that causes difficulties in training. Despite the good performance in computer vision, GAN based methods do not sound real \cite{autovc}, as the discriminator is easier to fool than human ears. In many-to-many VC task, the quality of converted voices are degraded as more speakers are trained simultaneously \cite{dis_VAE}. 

In another research direction, variational  auto-encoders (VAE) based methods have a simple objective function, i.e. the maximization of ELBO (evidence lower bound) and its training strategy is suitable for self-supervised learning. 
Recent works \cite{autovc, autopst, dis_VAE} use autoencoder frameworks to disentangle input utterances into two embeddings which correspond to speaker and linguistic content information respectively. To generate a specific utterance from a target speaker, we use the concatenation of the speaker embedding of the target speaker and the content embedding of source speaker uttering the desired sentences. 
Previous works have showed that the   transformer structure has advantages in speech recognition  \cite{attn_vs_rnn_asr} and general speech applications compared to RNN's  \cite{attn_vs_rnn_speech}.  
In speech tasks, attention mechanism was first adopted for speech recognition \cite{las_asr} and  afterward used in VC tasks \cite{attn_vc_1,attn_vc_2,attn_vc_3,attn_vc_4,attn_vc_5}. Prior works mainly used cross-attention between source and target in latent space. 

Noticing that the speaker information is global and has long range dependency, we add self-attention to gain a more effective many-to-many zero-shot style transfer, i.e. attention disentangled VAE. We find that {\it self-attention VAE with structured pruning has considerable advantages in voice conversion especially on unseen speakers}. Merely adding self-attention to VAE is much less dramatic.

In our experimental section, we compared our method with Disentangled-VAE \cite{dis_VAE}\footnote{The baseline code miscalculated the KL divergence term, see github discussion of  \href{https://github.com/v-manhlt3/Disentangle-VAE-for-VC/issues/5}{this issue}. We adopted the corrected version.} and another recent method Fragment VC \cite{FragVC} on VCTK Corpus  \cite{vctk}. We employ group-$\ell_0$ penalty and a splitting algorithm \cite{rgsm} in our implementation to remarkably improve model generalization on unseen speakers. Our converted utterances out-perform  \cite{dis_VAE,FragVC} in both voice quality and conversion accuracy measured objectively by third party packages \cite{mosnet, resemblyzer}.

\section{Related Works}
\subsection{Zero Shot Voice Conversion}
IDE-VC \cite{yuan2021improving} learned disentangled representation by introducing coarse lower/upper bounds of mutual information. 
Sequential AutoEncoder \cite{seq_vae, s3vae} replaced standard normal prior with time-dependent learnable prior.
Later works \cite{dis_VAE, adainvc, autovc, yuan2021improving} improved the performance of disentangled VAE to enable zero-shot voice conversion, i.e. the network has never listened to the voice of source/target speaker. Disentangled-VAE \cite{dis_VAE} used $\beta$-VAE \cite{beta_vae}, which modified the variational ELBO of VAE to encourage disentangled representations, and learn separate features for speaker and content respectively. 
AutoPST \cite{autopst} furthermore disentangled prosody by re-sampling, which captures speaker's rhythm information. 

\subsection{Attention Based Voice Conversion}
ATTS2S-VC \cite{attn_vc_1} is a  sequence-to-sequence model trained on parallel data with RNN appearing in  encoder and decoder, and with guided cross-attention layer to do conversion in latent space. FastS2S-VC \cite{attn_vc_2} used the same idea and improved the inference speed to real time voice conversion. Similarly, \cite{attn_vc_3} also used cross-attention in latent space, but the encoder and decoder used CNN instead of RNN and one-shot voice conversion was supported. Apart from latent space, \cite{attn_vc_4,attn_vc_5} also tried attention to extract speaker identity information. FragmentVC \cite{FragVC} 
obtained the source utterance latent phonetic structure from Wav2Vec 2.0 and target 
utterance spectral features from 
log-mel-spectrograms. Aligning the hidden structures of the two different feature spaces with a two-stage training process,
FragmentVC extracts fine-grained voice fragments from target speaker utterance(s) 
and fuse them into the desired utterance.
Both self and cross attention mechanisms 
exist in the decoder. Different from previous works, {\it only self-attention is present in our VAE decoder to capture long-range dependency while keeping the model size increase due to adding attention moderate} (around 10\%). Moreover, we performed {\it structured pruning on the resulting VAE to reduce its over-fitting and generalization error}. 

\begin{figure}[htb]
\begin{subfigure}[b]{\linewidth}
    \centering
    \includegraphics[width=\textwidth]{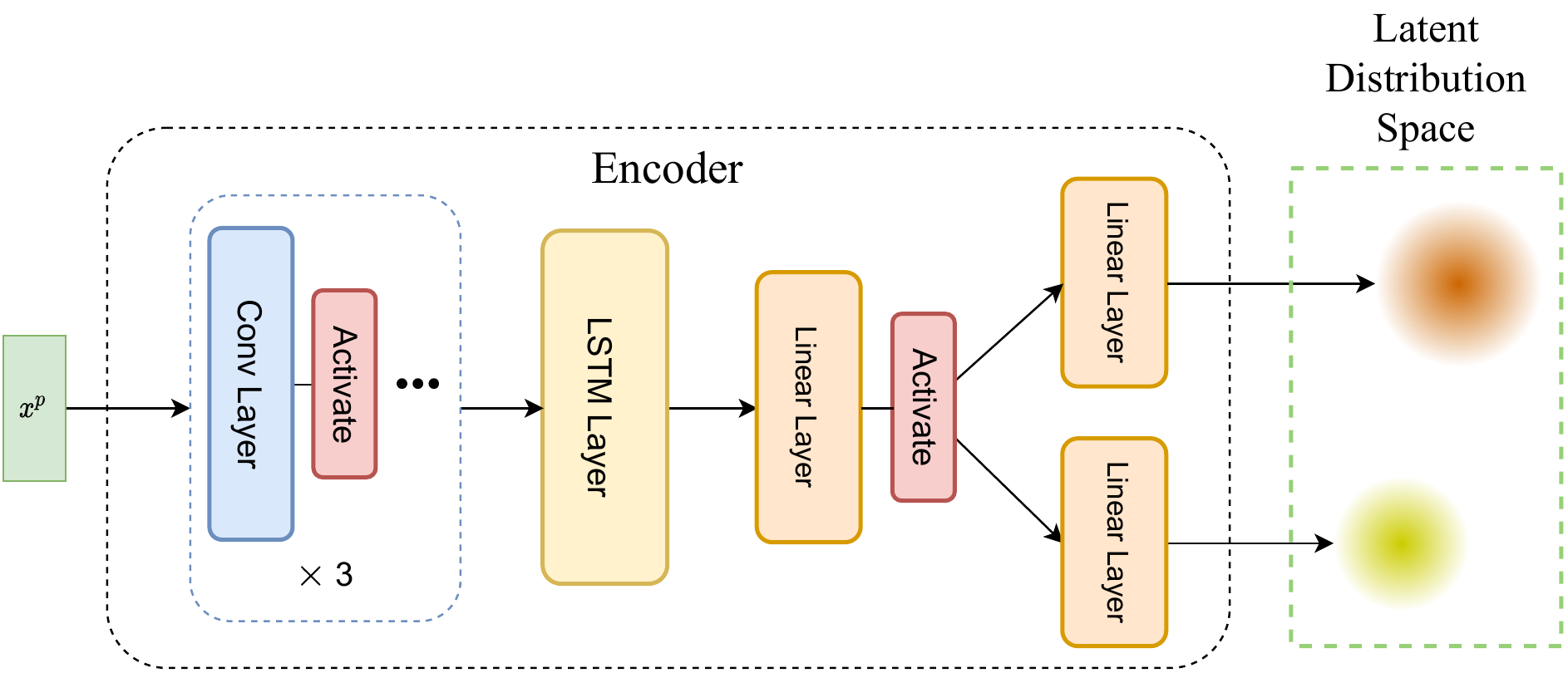}
    \caption{Neural net architecture of Encoder, $\times 3$ indicates the same network structure in dashed block repeated $3$ times.}
    \label{fig:model_en}
\end{subfigure}%
\vspace{0.5cm}
\begin{subfigure}[b]{\linewidth}
    \centering
    \includegraphics[width=\textwidth]{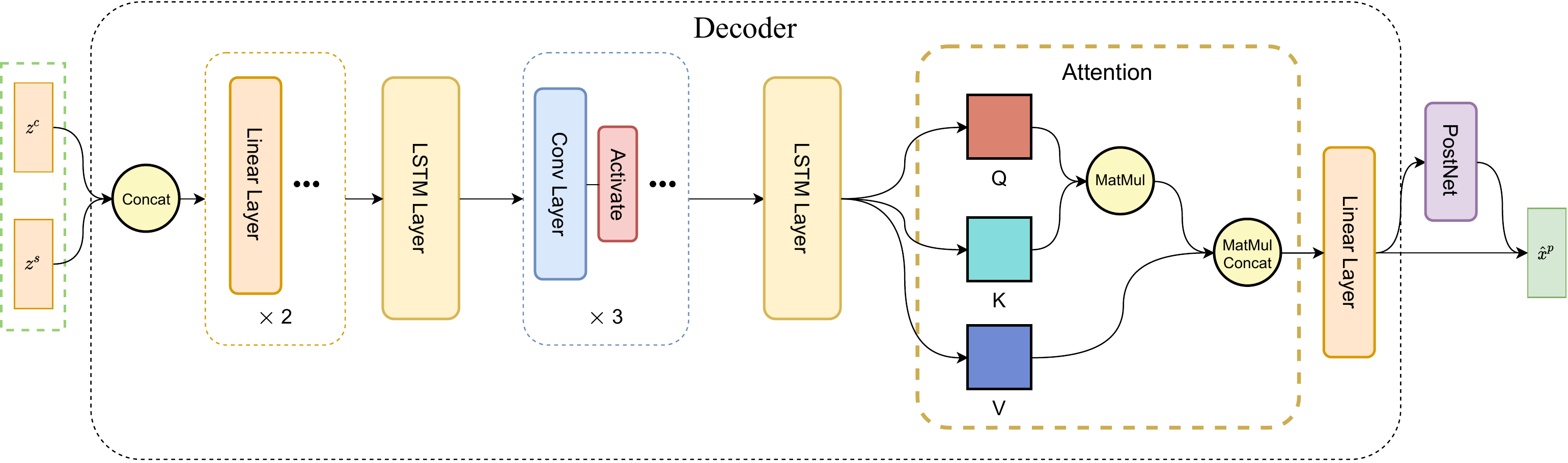}
    \caption{Neural net architecture of Decoder, with self-attention mechanism, $\times n$ indicates the same network structure in dashed block repeated $n$ times. }
    \label{fig:model_de}
\end{subfigure}%
\vspace{0.1cm}
\caption{Network structure of variational auto-encoder.}
\end{figure}

\begin{figure}[htb]
\begin{subfigure}{\linewidth}
    \centering
    \includegraphics[width=\linewidth]{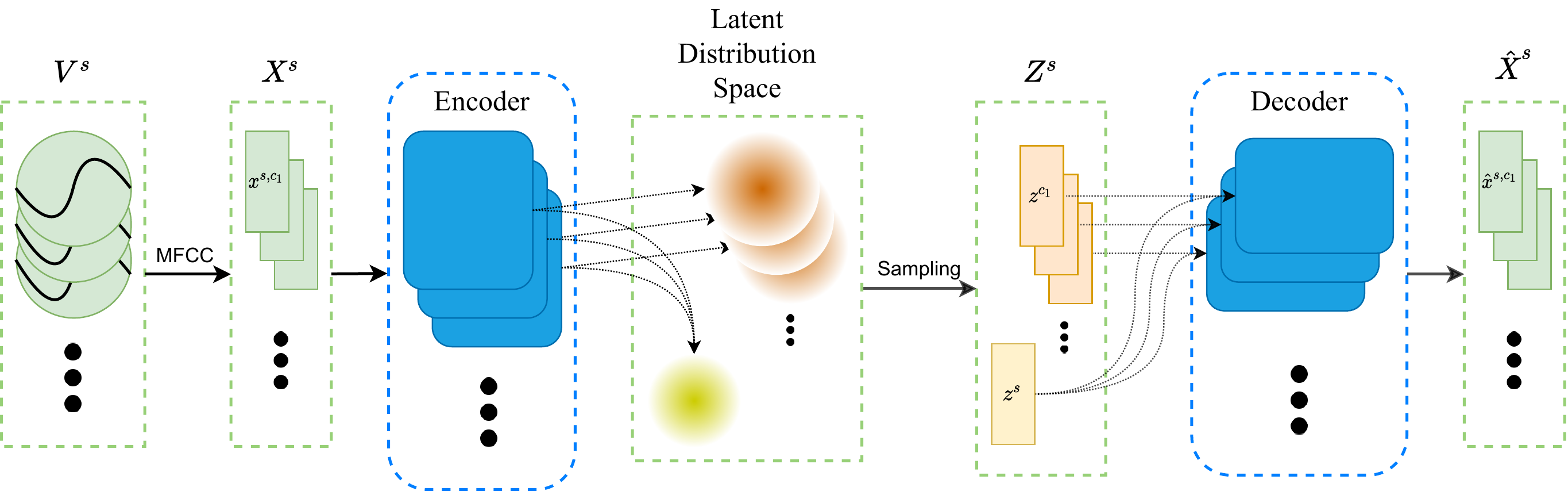}
    \caption{Model training process for one speaker.}
    \label{fig:model_tr}
\end{subfigure}%
\vspace{0.5cm}
\begin{subfigure}{\linewidth}
    \centering
    \includegraphics[width=\linewidth]{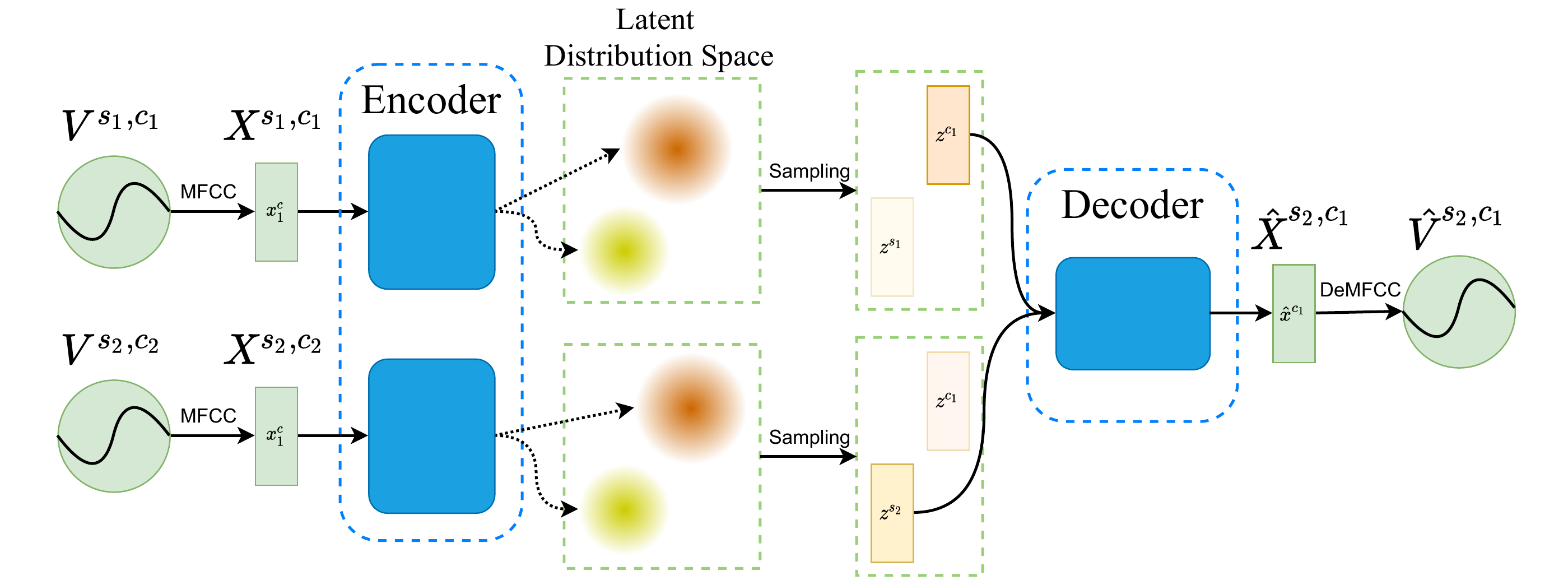}
    \caption{An example of voice conversion process of generating voice $\hat{V}^{s_2, c_1}$ with speaker $s_2$ and linguistic content $c_1$.}
    \label{fig:model_ed}
\end{subfigure}%
\vspace{0.1cm}
\caption{Model training and voice conversion process of our proposed method.}
\end{figure}

\section{Proposed Approach}
\subsection{Preliminary}
VAE \cite{vae} is an autoencoder that compresses the input into a regularized latent distribution in the encoder, and reconstructs the input back in the decoder. More precisely, let the encoder extract posterior distribution $q_\phi(\z|\x)$ of a group of latent variables $\z$ given an input data $\x$, while the decoder recovers the conditional distribution $p_{\theta}(\x|\z)$ of input data $\x$ given a sample of the random variable $\z$. Since the marginal likelihood $p_\theta(\x)=\int p(\z) p_\theta(\x|\z)d\z$ is intractable for a large dataset, it is hard to maximize directly. A variational lower bound(ELBO) is often optimized instead:
$$\mathbb{E}_{q_\phi(\z|\x)}\left[\log p_\theta(\x|\z)\right]-D_{KL}\left(q_\phi(\z|\x)||p(\z)\right),$$
which is the same as minimizing the sum of the reconstruction error and Kullaback-Leibler divergence between the posterior and the prior. The $\beta$-VAE \cite{beta_vae} is a modification of VAE with an emphasis on discovering disentangled latent factors. The objective function to maximize, different from the original, is 
$$\mathbb{E}_{q_\phi(\z|\x)}\left[\log p_\theta(\x|\z)\right]-\beta D_{KL}\left(q_\phi(\z|\x)||p(\z)\right),$$
where $\beta>1$ is a trade-off between reconstruction quality and the extent of disentanglement. 

\subsection{Problem Formulation}
We build on network structure and objective function in \cite{beta_vae, dis_VAE} by adding a suitable attention mechanism \cite{attention} to the decoder and train with Group-$\ell_0$ Splitting Method (RGSM, \cite{rgsm}) to alleviate over-fitting. 
Denote by $\bm v^{s_i,c_i}$ for speaker $s_i$ to utter the linguistic content $c_i$, and by  $\x^{s_i,c_i}$ the 80 dimensional log-MFCC of the corresponding utterance. The goal of this work is to generate $\bm v^{s_j, c_i}$ from $\bm v^{s_i, c_i}$ and $\bm v^{s_j, c_j}$ where $s_i$ and $s_j$ could be unseen during training, and $i\not = j$. 
\subsection{Encoder}
We assume that the latent space $\z$ of any utterance $\x$ is the direct sum of two sub-latent spaces $\z_s$ and $\z_c$ that correspond to the speaker $s$ and content $c$ of that utterance respectively, where the distribution of $\z:=(\z_s, \z_c)$ can be extracted by a well-designed encoder $\enc$. 

Same as the standard VAE \cite{vae}, we assume that $\z$ is normally distributed with a diagonal co-variance matrix and use re-parameterization trick to sample $\z$ during the training stage. In short, we assume that
$\left(\z_s,\z_c\right)\sim p_\phi(\z_s,\z_c|\x)=\enc(\x;\phi)$,
where $\z_s$ and $\z_c$ are assumed to be normally  distributed $\z_s\sim \mathcal{N}(\m_s, \si_s)$ and $\z_s\sim \mathcal{N}(\m_c, \si_c)$\footnote{Since we consider only Gaussians with diagonal co-variances, we specify a vector of variances in the second parameter to Gaussian distribution as convention.} respectively. 

\subsection{Decoder}
Since we have assumed that the speaker's sub-latent representation does not change much between different utterances of the same speaker, we adopt group based learned representation \cite{group_based}, where the decoder takes each content embedding $\z_c$ sampled from distribution of corresponding utterance and one speaker embedding $\z_s$ sampled from the ``average distribution'' of all utterances in the same group, i.e.
$$\z_s^{(i)}\sim\mathcal{N}\left(\frac{1}{n}\sum_{j=1}^n\m_s^{(j)},\left(\mathop{\bigodot}_{j=1}^n\si_s^{(j)}\right)^{1/n}\right)$$
for $i\in[n]$ where $n$ is the group size. Decoder gives reconstructed input as follows:
$$\hat{\x}=\dec\left(\z|\theta\right)=\dec\left(\z_s,\z_c|\theta\right).$$

\subsection{Objective Function}
Same as in \cite{dis_VAE}, we also adopt Post-Net \cite{tacotron2}, namely $\post$, to refine the reconstructed log-MFCC so that our reconstruction loss splits into two terms:
$$\mathcal{L}_{rec}=\norm{\hat{\x}-\x}+\norm{\hat{\x}+\post\left(\hat{\x}\right)-\x}$$
where $\hat{\x}=\dec\left(\z|\theta\right)$ is the decoder output.

Our loss function is a variant of \textbf{negative} Evidence Lower Bound (ELBO): 
$$\mathcal{L}=\mathcal{L}_{rec}+\beta\ D_{KL}\left(p_{\phi}\left(\z|\x\right)||\mathcal{N}(\bm 0, \bm I)\right),$$
for some $\beta\geq1$ which was proposed in \cite{beta_vae}.

\subsection{Relaxed Group-wise Splitting Method}

Consider $\bm W=\{ \cdots, w_g, \cdots \}$, $1 \leq g \leq G$ as grouped weights of a layer in VAE model, where $G$ is the number of groups. The group $\ell_0$ (G$l_0$) penalty is: $ ||\bm W||_{Gl_0} := \Sigma^{G}_{g=1} \bm 1_{||w_g||_2 \neq 0} $, where $\bm 1$ is the  characteristic function, and its standard convex relaxation the   
group Lasso (GL) penalty \cite{yuan2006model} is: $ ||\bm W||_{GL} := \Sigma^{G}_{g=1} ||w_g||_2 $.
By solving exactly the following G$\ell_0$ proximal problem for positive parameter $\lambda$:
$$    \underset{z_g}{\arg \min} \ \lambda \, \bm 1_{\|z_g\|\neq 0} + \frac{1}{2}\, ||z_g-w_g||_2 $$
we deduce the proximal ({\it projection} or {\it shrinkage}) operator of G$\ell_0$ penalty: 
$$
 \textbf{Prox}_{Gl_0, \lambda} (w_g) := w_g \bm 1 _{||w_g||_2> \sqrt{2\lambda}}.
$$
Similarly, 
$ \textbf{Prox}_{GL, \lambda} (w_g) := w_g \max\{ ||w_g||_2-\lambda, 0 \} / ||w_g||_2. 
$
Both operators are group-wise operations.
Applying group-wise splitting and proximal operators to gradient descent training, we implemented the relaxed group-wise splitting method (RGSM) \cite{rgsm,Dinh:LOD} 
summarized in Alg.\ref{alg:rgsm}, where $\alpha$ is the learning rate of gradient descent, $\beta$ is a positive (relaxation) parameter, $\delta = G\ell_0\ or \ GL$ refers to the type of penalty. 

In our work here on voice conversion, RGSM is applied to {\it each VAE layer having width greater than $128$, which turns out to be either a fully-connected layer or an LSTM layer} where {\it group means column of a weight matrix}. In our experiments, the group $\ell_0$ penalty ($\delta=G\ell_0$) is found better than group Lasso and is adopted for all results reported below. 
Because the {\it group $\ell_0$ penalty is discontinuous}, the splitting step in RGSM is absolutely necessary for the penalty to be integrated into the stochastic gradient descent training.
The hyper-parameters we chose for RGSM are $\beta=0.1$, $\lambda=4\times10^{-2}$, $\lambda_l=10^{-6}$.
Due to its {\it built-in shrinkage mechanism, RGSM helps to reduce overfitting in network training and considerably improve the generalization capability} of our proposed self-attention VAE model as we shall see in section IV.

\begin{algorithm}
\caption{Relaxed Group-wise Splitting Method}\label{alg:rgsm}
\begin{algorithmic}[1]
\STATE Hyper-parameters $\lambda, \lambda_l, \beta, \alpha$
\STATE Objective function: $f(w) = \text{loss}(w, x) + \lambda_l||w||_{GL}$
\STATE Initialize $w^0$
\FOR{$e = 1, \cdots, \text{max-epoch},$}
    \FOR{$g=1,\cdots,G,$}
    \STATE $u^{e} = \text{Prox}_{\delta, \lambda} (w_{e-1})$;
    \ENDFOR
    \STATE $ w^{e+1} = w^{e} - \alpha \nabla f(w) -\alpha \beta(w^{e}-u^{e}) $;
\ENDFOR
\end{algorithmic}
\end{algorithm}

\section{Experiments}
\subsection{Dataset and experimental setting}
\begin{table}[h]
\begin{center}
\begin{tabular}{| *{3}{c|} }\hline
Block    & \multicolumn{2}{c|}{Layer}\\ \hline
\multirow{6}{*}{Encoder} & Input layer & $64\times80$ \\ \cline{2-3}
& 1D Conv layer $\times3$ & (5, 2, 512) \\ \cline{2-3}
& BiLSTM layer & (512, 64, 2) \\ \cline{2-3}
& FC layer & (8192, 2048) \\ \cline{2-3}
& FC layer (content) & (2048, 56) \\ \cline{2-3}
& FC layer (speaker)& (256, 8) \\ \hline
\multirow{7}{*}{Decoder} & FC layer & (32, 2048) \\ \cline{2-3}
& FC layer & (2048, 8192) \\ \cline{2-3}
& LSTM layer & (128, 512, 1) \\ \cline{2-3}
& 1D Conv Layer $\times3$ & (5,2,512) \\ \cline{2-3}
& LSTM layer & (512, 1024, 2) \\ \cline{2-3}
& \textbf{Multihead Self-Attention} & (8, 128) \\ \cline{2-3}
& FC layer& (1024, 80) \\
\hline
\end{tabular}
\end{center}
\caption{The architecture of our model. The parameters of the 1D convolution layer are denoted as kernel size, stride, output channel number. The parameters of the LSTM layer are input size, hidden layer width, and the number of hidden layers. The parameters of the multihead attention are number of head and feature dimension of each head.}
\label{tab:network}
\end{table}

We use VCTK Corpus \cite{vctk} which includes hundreds of utterances of 109 speakers respectively. We split the data set into two parts: a training set and a testing set. The test set contains all utterances of 6 speakers (3 males and 3 females) and 10 utterances of the rest speakers. The six unseen speakers are p225 (female), p226 (male), p227 (male), p228 (female), p229 (female), and p232 (male), while the seen speakers to evaluate are p286 (male), p287 (male) p288 (female), p292 (male), p293 (female) and p294 (female). The feature we fed to and the output of our network are both 80 dimensional log MFCC (Mel-frequency cepstral coefficients), for which the STFT (short-term Fourier transform) is performed with window size 1024 and hop length 256. We chose the vocoder from wavnet \cite{wavnet}. The Table \ref{tab:network} presents the detail of our model parameter settings.

\subsection{Evaluation}
To evaluate our method, we consider two objective metrics: i) voice quality and ii) speaker classification accuracy. Both metrics are measured with third-party pre-trained network and applied to conversion between both seen and unseen speakers. We picked 6 seen and 6 unseen speakers and do 10 cross utterance conversions for each pair. 

For each converted utterance, the voice quality is measured by MOSNet \cite{mosnet} that predicts human ratings of converted speech. MOSnet score ranges from 1 to 5, with lowest score of 1 and highest score of 5. 
Table \ref{mos_score} compares MOSNET scores of converted utternance of seen/unseen speakers from VAE models with/without attention and RGSM in training, showing that {\it the objective speech quality is improved by having RGSM}.

\begin{table}[h]
\begin{center}
\begin{tabular}{|c|c|c|}\hline
                           & seen  & unseen \\ \hline
 Baseline                  & 3.59  & 3.35   \\
 FragmentVC                & 3.01  & 3.07   \\ 
 Attention                 & 3.49  & 3.39   \\ 
 RGSM                      & \textbf{3.74}  & \textbf{3.58}   \\
 Attention + RGSM          & 3.63  & 3.51   \\ \hline
\end{tabular}
\end{center}
\caption{Average MOSNet scores of converted utternance of seen/unseen speaker and with/without attention.}
\label{mos_score}
\end{table}

Given any audio file of a speech, Resemblyzer \cite{resemblyzer} creates a summary vector of 256 values summarizing the characteristics of the voice spoken. Given multiple speech wav files of a speaker, the Resemblyzer speaker embedding is the mean of multiple Resemblyzer speech embeddings. From the full VCTK dataset \cite{vctk}, we generated six speaker embeddings of seen and unseen speakers respectively. For each converted utterance, we generate the utterance embedding and select the speaker with closest speaker embedding among the six candidates. The prediction is correct if the selected speaker is the target speaker. Since our classifier picks the closest embedding among the six known speaker embeddings, no threshold is involved in our classification. Table \ref{speaker_acc} compares the average speaker classification accuracy of converted utterances of seen/unseen speakers with/without attention
and with/without RGSM in training, showing that VAE model with attention layer and trained by RGSM reaches the highest classification accuracy. 

\begin{table}[h]
\begin{center}
\begin{tabular}{|c|c|c|}\hline
                           & seen    & unseen \\ \hline
 Baseline                  & 78.6\%  & 36.7\%   \\
 FragmentVC                & 67.3\%  & 63.3\%   \\ 
 Attention                 & 84.0\%  & 37.7\%   \\ 
 RGSM                      & 80.0\%  & 34.3\%   \\
 Attention + RGSM          & \textbf{93.3\%}  & \textbf{65.0\%}   \\ \hline
\end{tabular}
\end{center}
\caption{Average speaker classification accuracy of converted utterances of seen/unseen speakers and with/without attention.}
\label{speaker_acc}
\end{table}



\section{Conclusions}
We successfully integrated a self-attention layer into 
the decoder of a VAE framework to enhance the zero-shot many-to-many
voice conversion task.
To improve generalization due to the ensuing larger model capacity, an efficient 
group-wise splitting and thresholding algorithm 
has been found efficient in 
maintaining the generated voice quality of VAE while significantly increasing speaker classification accuracy of converted utterance of seen/unseen speakers. In future work, we plan to 
explore some of the feature extraction and attention 
structures in \cite{FragVC} to further reduce generalization error.

\bibliographystyle{IEEEtran}
\bibliography{main}

\begin{thebibliography}{10}
\providecommand{\url}[1]{#1}
\csname url@samestyle\endcsname
\providecommand{\newblock}{\relax}
\providecommand{\bibinfo}[2]{#2}
\providecommand{\BIBentrySTDinterwordspacing}{\spaceskip=0pt\relax}
\providecommand{\BIBentryALTinterwordstretchfactor}{4}
\providecommand{\BIBentryALTinterwordspacing}{\spaceskip=\fontdimen2\font plus
\BIBentryALTinterwordstretchfactor\fontdimen3\font minus
  \fontdimen4\font\relax}
\providecommand{\BIBforeignlanguage}[2]{{%
\expandafter\ifx\csname l@#1\endcsname\relax
\typeout{** WARNING: IEEEtran.bst: No hyphenation pattern has been}%
\typeout{** loaded for the language `#1'. Using the pattern for}%
\typeout{** the default language instead.}%
\else
\language=\csname l@#1\endcsname
\fi
#2}}
\providecommand{\BIBdecl}{\relax}
\BIBdecl

\bibitem{parrotron}
F.~Biadsy, R.~J. Weiss, P.~J. Moreno, D.~Kanevsky, and Y.~Jia, ``Parrotron: An
  end-to-end speech-to-speech conversion model and its applications to
  hearing-impaired speech and speech separation,'' 2019.

\bibitem{huang2019voice}
W.-C. Huang, T.~Hayashi, Y.-C. Wu, H.~Kameoka, and T.~Toda, ``Voice transformer
  network: Sequence-to-sequence voice conversion using transformer with
  text-to-speech pretraining,'' 2019.

\bibitem{lee2006map}
C.-H. Lee and C.-H. Wu, ``Map-based adaptation for speech conversion using
  adaptation data selection and non-parallel training,'' in \emph{Ninth
  International Conference on Spoken Language Processing}, 2006.

\bibitem{saito2011one}
D.~Saito, K.~Yamamoto, N.~Minematsu, and K.~Hirose, ``One-to-many voice
  conversion based on tensor representation of speaker space,'' in
  \emph{Twelfth Annual Conference of the International Speech Communication
  Association}, 2011.

\bibitem{xie2016kl}
F.-L. Xie, F.~K. Soong, and H.~Li, ``A {KL} divergence and {DNN}-based approach
  to voice conversion without parallel training sentences.'' in
  \emph{Interspeech}, 2016, pp. 287--291.

\bibitem{hsu2017voice}
C.-C. Hsu, H.-T. Hwang, Y.-C. Wu, Y.~Tsao, and H.-M. Wang, ``Voice conversion
  from unaligned corpora using variational autoencoding wasserstein generative
  adversarial networks,'' \emph{arXiv preprint arXiv:1704.00849}, 2017.

\bibitem{kaneko2017parallel}
T.~Kaneko and H.~Kameoka, ``Parallel-data-free voice conversion using
  cycle-consistent adversarial networks,'' \emph{arXiv preprint
  arXiv:1711.11293}, 2017.

\bibitem{kameoka2019acvae}
H.~Kameoka, T.~Kaneko, K.~Tanaka, and N.~Hojo, ``Acvae-vc: Non-parallel voice
  conversion with auxiliary classifier variational autoencoder,''
  \emph{IEEE/ACM Transactions on Audio, Speech, and Language Processing},
  vol.~27, no.~9, pp. 1432--1443, 2019.

\bibitem{kaneko2018cyclegan}
T.~Kaneko and H.~Kameoka, ``Cyclegan-vc: Non-parallel voice conversion using
  cycle-consistent adversarial networks,'' in \emph{2018 26th European Signal
  Processing Conference (EUSIPCO)}.\hskip 1em plus 0.5em minus 0.4em\relax
  IEEE, 2018, pp. 2100--2104.

\bibitem{kaneko2020cyclegan}
T.~Kaneko, H.~Kameoka, K.~Tanaka, and N.~Hojo, ``Cyclegan-vc3: Examining and
  improving cyclegan-vcs for mel-spectrogram conversion,'' \emph{arXiv preprint
  arXiv:2010.11672}, 2020.

\bibitem{kameoka2018stargan}
H.~Kameoka, T.~Kaneko, K.~Tanaka, and N.~Hojo, ``Stargan-vc: Non-parallel
  many-to-many voice conversion using star generative adversarial networks,''
  in \emph{2018 IEEE Spoken Language Technology Workshop (SLT)}.\hskip 1em plus
  0.5em minus 0.4em\relax IEEE, 2018, pp. 266--273.

\bibitem{kaneko2019stargan}
T.~Kaneko, H.~Kameoka, K.~Tanaka, and N.~Hojo, ``Stargan-vc2: Rethinking
  conditional methods for stargan-based voice conversion,'' \emph{arXiv
  preprint arXiv:1907.12279}, 2019.

\bibitem{zhang2020gazev}
Z.~Zhang, B.~He, and Z.~Zhang, ``Gazev: Gan-based zero-shot voice conversion
  over non-parallel speech corpus,'' \emph{arXiv preprint arXiv:2010.12788},
  2020.

\bibitem{choi2018stargan}
Y.~Choi, M.~Choi, M.~Kim, J.-W. Ha, S.~Kim, and J.~Choo, ``Stargan: Unified
  generative adversarial networks for multi-domain image-to-image
  translation,'' in \emph{Proceedings of the IEEE conference on computer vision
  and pattern recognition}, 2018, pp. 8789--8797.

\bibitem{autovc}
K.~Qian, Y.~Zhang, S.~Chang, X.~Yang, and M.~Hasegawa-Johnson, ``Autovc:
  Zero-shot voice style transfer with only autoencoder loss,'' 2019.

\bibitem{dis_VAE}
M.~Luong and V.~A. Tran, ``Many-to-many voice conversion based feature
  disentanglement using variational autoencoder,'' 2021.

\bibitem{autopst}
K.~Qian, Y.~Zhang, S.~Chang, J.~Xiong, C.~Gan, D.~Cox, and M.~Hasegawa-Johnson,
  ``Global rhythm style transfer without text transcriptions,'' 2021.

\bibitem{attn_vs_rnn_asr}
A.~Zeyer, P.~Bahar, K.~Irie, R.~Schlüter, and H.~Ney, ``A comparison of
  transformer and lstm encoder decoder models for asr,'' in \emph{2019 IEEE
  Automatic Speech Recognition and Understanding Workshop (ASRU)}, 2019, pp.
  8--15.

\bibitem{attn_vs_rnn_speech}
S.~Karita, N.~Chen, T.~Hayashi, T.~Hori, H.~Inaguma, Z.~Jiang, M.~Someki,
  N.~E.~Y. Soplin, R.~Yamamoto, X.~Wang, S.~Watanabe, T.~Yoshimura, and
  W.~Zhang, ``A comparative study on transformer vs rnn in speech
  applications,'' in \emph{2019 IEEE Automatic Speech Recognition and
  Understanding Workshop (ASRU)}, 2019, pp. 449--456.

\bibitem{las_asr}
W.~Chan, N.~Jaitly, Q.~Le, and O.~Vinyals, ``Listen, attend and spell: A neural
  network for large vocabulary conversational speech recognition,'' in
  \emph{2016 IEEE International Conference on Acoustics, Speech and Signal
  Processing (ICASSP)}, 2016, pp. 4960--4964.

\bibitem{attn_vc_1}
K.~Tanaka, H.~Kameoka, T.~Kaneko, and N.~Hojo, ``Atts2s-vc:
  Sequence-to-sequence voice conversion with attention and context preservation
  mechanisms,'' 11 2018.

\bibitem{attn_vc_2}
\BIBentryALTinterwordspacing
H.~Kameoka, K.~Tanaka, and T.~Kaneko, ``Fasts2s-vc: Streaming
  non-autoregressive sequence-to-sequence voice conversion,'' \emph{CoRR}, vol.
  abs/2104.06900, 2021. [Online]. Available:
  \url{https://arxiv.org/abs/2104.06900}
\BIBentrySTDinterwordspacing

\bibitem{attn_vc_3}
T.~Ishihara and D.~Saito, ``Attention-based speaker embeddings for one-shot
  voice conversion,'' 10 2020, pp. 806--810.

\bibitem{attn_vc_4}
Y.~Zhang, H.~Che, J.~Li, C.~Li, X.~Wang, and Z.~Wang, ``One-shot voice
  conversion based on speaker aware module,'' in \emph{ICASSP 2021 - 2021 IEEE
  International Conference on Acoustics, Speech and Signal Processing
  (ICASSP)}, 2021, pp. 5959--5963.

\bibitem{attn_vc_5}
S.~Liu, J.~Zhong, L.~Sun, X.~Wu, X.~Liu, and H.~Meng, ``Voice conversion across
  arbitrary speakers based on a single target-speaker utterance,'' 09 2018, pp.
  496--500.

\bibitem{FragVC}
Y.~Lin, C.-M. Chien, J.-H. Lin, H.-Y. Lee, and L.-S. Lee, ``Fragmentvc:
  Any-to-any voice conversion by end-to-end extracting and fusing fine-grained
  voice fragments with attention,'' in \emph{IEEE International Conference on
  Acoustics, Speech and Signal Processing (ICASSP)}, 2021, pp. 5939--5943.

\bibitem{vctk}
\BIBentryALTinterwordspacing
C.~Veaux and a.~K.~M. Junichi~Yamagishi, ``\BIBforeignlanguage{en}{Superseded -
  {CSTR} {VCTK} corpus: English multi-speaker corpus for cstr voice cloning
  toolkit},'' 2017. [Online]. Available:
  \url{https://datashare.ed.ac.uk/handle/10283/2651}
\BIBentrySTDinterwordspacing

\bibitem{rgsm}
B.~Yang, J.~Lyu, S.~Zhang, Y.-Y. Qi, and J.~Xin, ``Channel pruning for deep
  neural networks via a relaxed group-wise splitting method,'' \emph{In Proc.
  of International Conference on AI for Industries, Laguna Hills, CA}, 2019.

\bibitem{mosnet}
C.-C. Lo, S.-W. Fu, W.-C. Huang, X.~Wang, J.~Yamagishi, Y.~Tsao, and H.-M.
  Wang, ``Mosnet: Deep learning based objective assessment for voice
  conversion,'' 2019.

\bibitem{resemblyzer}
G.~Louppe, ``Resemblyzer,'' \url{https://github.com/resemble-ai/Resemblyzer},
  2019.

\bibitem{yuan2021improving}
\BIBentryALTinterwordspacing
S.~Yuan, P.~Cheng, R.~Zhang, W.~Hao, Z.~Gan, and L.~Carin, ``Improving
  zero-shot voice style transfer via disentangled representation learning,'' in
  \emph{International Conference on Learning Representations}, 2021. [Online].
  Available: \url{https://openreview.net/forum?id=TgSVWXw22FQ}
\BIBentrySTDinterwordspacing

\bibitem{seq_vae}
Y.~Li and S.~Mandt, ``Disentangled sequential autoencoder,'' 2018.

\bibitem{s3vae}
Y.~Zhu, M.~R. Min, A.~Kadav, and H.~P. Graf, ``S3vae: Self-supervised
  sequential {VAE} for representation disentanglement and data generation,''
  2020.

\bibitem{adainvc}
J.~Chou, C.~Yeh, and H.~Lee, ``One-shot voice conversion by separating speaker
  and content representations with instance normalization,'' 2019.

\bibitem{beta_vae}
\BIBentryALTinterwordspacing
I.~Higgins, L.~Matthey, A.~Pal, C.~Burgess, X.~Glorot, M.~Botvinick,
  S.~Mohamed, and A.~Lerchner, ``$\beta$-vae: Learning basic visual concepts
  with a constrained variational framework,'' in \emph{International Conference
  on Learning Representations}, 2017. [Online]. Available:
  \url{https://openreview.net/forum?id=Sy2fzU9gl}
\BIBentrySTDinterwordspacing

\bibitem{vae}
D.~Kingma and M.~Welling, ``Auto-encoding variational {Bayes},'' 2014.

\bibitem{attention}
\BIBentryALTinterwordspacing
A.~Vaswani, N.~Shazeer, N.~Parmar, J.~Uszkoreit, L.~Jones, A.~N. Gomez, L.~u.
  Kaiser, and I.~Polosukhin, ``Attention is all you need,'' in \emph{Advances
  in Neural Information Processing Systems}, I.~Guyon, U.~V. Luxburg,
  S.~Bengio, H.~Wallach, R.~Fergus, S.~Vishwanathan, and R.~Garnett, Eds.,
  vol.~30.\hskip 1em plus 0.5em minus 0.4em\relax Curran Associates, Inc.,
  2017. [Online]. Available:
  \url{https://proceedings.neurips.cc/paper/2017/file/3f5ee243547dee91fbd053c1c4a845aa-Paper.pdf}
\BIBentrySTDinterwordspacing

\bibitem{group_based}
\BIBentryALTinterwordspacing
H.~Hosoya, ``A simple probabilistic deep generative model for learning
  generalizable disentangled representations from grouped data,'' \emph{CoRR},
  vol. abs/1809.02383, 2018. [Online]. Available:
  \url{http://arxiv.org/abs/1809.02383}
\BIBentrySTDinterwordspacing

\bibitem{tacotron2}
\BIBentryALTinterwordspacing
J.~Shen, R.~Pang, R.~J. Weiss, M.~Schuster, N.~Jaitly, Z.~Yang, Z.~Chen,
  Y.~Zhang, Y.~Wang, R.~J. Skerry{-}Ryan, R.~A. Saurous, Y.~Agiomyrgiannakis,
  and Y.~Wu, ``Natural {TTS} synthesis by conditioning wavenet on mel
  spectrogram predictions,'' \emph{CoRR}, vol. abs/1712.05884, 2017. [Online].
  Available: \url{http://arxiv.org/abs/1712.05884}
\BIBentrySTDinterwordspacing

\bibitem{yuan2006model}
M.~Yuan and Y.~Lin, ``Model selection and estimation in regression with grouped
  variables,'' \emph{Journal of the Royal Statistical Society: Series B
  (Statistical Methodology)}, vol.~68, no.~1, pp. 49--67, 2006.

\bibitem{Dinh:LOD}
T.~Dinh, B.~Wang, A.~Bertozzi, S.~Osher, and J.~Xin, ``Sparsity meets
  robustness: Channel pruning for the {F}eynman-{K}ac formalism principled
  robust deep neural nets,'' in \emph{Proc. of International Conference on
  Machine Learning, Optimization, and Data Science; and LNCS}, vol. 12566,
  2020, pp. 362--381.

\bibitem{wavnet}
\BIBentryALTinterwordspacing
A.~van~den Oord, S.~Dieleman, H.~Zen, K.~Simonyan, O.~Vinyals, A.~Graves,
  N.~Kalchbrenner, A.~W. Senior, and K.~Kavukcuoglu, ``Wavenet: {A} generative
  model for raw audio,'' \emph{CoRR}, vol. abs/1609.03499, 2016. [Online].
  Available: \url{http://arxiv.org/abs/1609.03499}
\BIBentrySTDinterwordspacing

\end{thebibliography}
\end{document}